\documentclass[aps,pre,superscriptaddress,amsmath,amssymb,twocolumn,showpacs,floatfix]{revtex4-1}

\usepackage{bm}% bold math
\usepackage{amssymb}
\usepackage{colordvi}
\usepackage{color}     
\usepackage{amsbsy}
\usepackage{amsmath}
\usepackage{amsbsy}
\usepackage{latexsym}
\usepackage[pdftex]{graphicx}

\newcommand{\OO}{O}            % overlap
\newcommand{\ave}[1]{\langle #1 \rangle}

\newcommand{\ETHmat}{ETH Z\"urich, Department of Materials, Polymer
Physics, HCI H541, CH-8093 Z\"urich, Switzerland}
\newcommand{\ETHbau}{ETH Z\"urich, Department of Civil Engineering, 
Institute for Building Materials, CH-8093 Z\"urich, Switzerland}

\begin{document}
\title{
Soft modes and non-affine rearrangements in the inherent structures of supercooled liquids
}

\author{Majid Mosayebi}
\altaffiliation{Present address: Physical and Theoretical Chemistry Laboratory, University of Oxford, Oxford, OX1 3QZ, United Kingdom}
\affiliation{\ETHmat}
\author{Patrick Ilg}
\affiliation{\ETHmat}
\author{Asaph Widmer-Cooper}
\affiliation{School of Chemistry, University of Sydney, NSW 2006, Australia}
\author{Emanuela Del Gado}
\affiliation{\ETHbau}
%\date{\today}

\begin{abstract}
We find that the hierarchical organization of the potential energy landscape in a model supercooled liquid can be related to a change in the spatial distribution of soft normal modes. 
For groups of nearby minima, between which fast relaxation processes typically occur, the localization of the soft modes 
is very similar. The spatial distribution of soft regions changes, instead, for minima between which transitions relevant to structural relaxation occur. This may be the reason why the soft modes are able to predict spatial heterogeneities in the 
dynamics. Nevertheless, the very softest modes are only weakly correlated with dynamical heterogeneities, and instead show higher statistical overlap with regions in the local minima that would undergo non-affine rearrangements if subjected to a shear deformation. This feature of the supercooled liquid is reminiscent of the behavior of non-affine deformations in amorphous solids, where the very softest modes identify the {\it loci} of plastic instabilities.   

\end{abstract}


\maketitle

Predicting the dynamical and mechanical behavior of simple liquids and crystalline solids from their structural properties can be successfully addressed by kinetic theories and elasticity, but this is not the case for glasses. Structurally, glasses and amorphous solids can be considered quite similar to supercooled liquids, but identifying a common link in their theoretical description is a major challenge. In both cases, the underlying potential energy landscape (PEL) is characterized by low-frequency normal modes that are quasi-localized in space \cite{Goldstein,Stillinger95,Heuer_ISreview,schober, harrowell_2012}. These soft modes represent the flattest directions near minima in the PEL and appear to play a primary role in both the dynamics of supercooled liquids and the mechanics of amorphous solids \cite{asaphSoftModes, asaphSoftModes2, Hocky_normalmode, manning, schall2012, wyart, Lemaitre06, Tsamados, anne_review,wyart2013}. In supercooled liquids the soft modes are microscopically correlated with dynamical heterogeneities (DH) and irreversible rearrangements. Studies of model amorphous solids subjected to quasi-static shear have also shown that the lowest frequency normal modes can predict the spatial location of plastic events and extended shear bands in amorphous solids just prior to the event occurring \cite{Lemaitre06,lemaitre2007,karmakar2010d}. 
These findings raise the question of whether the properties of supercooled liquids and amorphous solids are controlled by similar features of the PEL. For example, could slowing down of the liquid dynamics and the presence of plastic instabilities in amorphous solids be controlled by the same soft modes? 

Here we investigate the spatial distribution of soft modes in a model supercooled liquid and find evidence that the time scale separation typical of glassy dynamics can be ascribed to a qualitative change in the spatial distribution of the structurally soft regions. Nearby minima, that should lie within the same meta-basin (MB, i.e.~clusters of minima grouped together by rapid reversible dynamic transitions between them), have very similar soft mode localization. Instead, the mode localization appears to change significantly between meta-basins. 
We sample nearby minima by applying a shear deformation to a local minimum, or inherent structure (IS), of the potential energy landscape and find that regions undergoing non-affine displacements between minima are also correlated with the spatially heterogeneous dynamics associated with the initial configuration. These non-affinely rearranging domains overlap strongly with the softest normal modes, in a manner similar to non-affine deformations in amorphous solids, where the softest modes have been shown to identify the {\it loci} of plastic instabilities. These findings suggest that the soft modes and non-affinely rearranging regions of the local minima represent the spatially resolved features of the PEL that link supercooled liquids and amorphous solids.

%----------------------------- METHODS ------------------------------------
We study a model fragile glass former, a two-dimensional equimolar binary mixture of soft disks interacting via the purely repulsive potential $\phi(r)=\epsilon(\sigma_{ab}/r)^{12}$, where $\sigma_{12}=1.2\sigma_{11}$ and $\sigma_{22}=1.4\sigma_{11}$ \cite{Harrowell99,asaphSoftModes,asaphSoftModes2}. We use 50 statistically independent samples of size ranging between $N=2000$ and $N=8000$ particles, equilibrated at temperatures spanning the supercooled regime of dynamics (between $T=2.0$ and $T=0.36$). All quantities are reported in reduced units, with the length- and energy-scale $\sigma_{11}$ and $\epsilon$, respectively. For $T < 0.36$ we could not reach full equilibration. 

We start by comparing the spatial localization of soft regions in nearby configurations. To identify the structurally soft regions of each initial equilibrated configuration $X=\{{{\bf r}_i}\}$ we use the iso-configurational DW-factor (DWF). This is defined for each particle $i$ as the variance of its position during a time interval $t_{DW}=10\tau$, which corresponds to the short-time $\beta$-relaxation regime at $T=0.4$, averaged over multiple runs originating from $X$.  
In supercooled liquids, this quantity has been shown to be spatially correlated with the soft mode localization in the IS $X^{\rm q}=\{{\bf r}_i^{\rm q}\}$ belonging to $X$\cite{asaphSoftModes, asaphSoftModes2}. Starting from $X$, we sample nearby minima by computing the IS $X^{\rm dq}=\{{\bf r}_i^{\rm dq}\}$ of the deformed configuration ${\bf r}_i^{\rm d}(\gamma)={\bf r}_i+\gamma y_i{\bf e}_x$, using conjugate gradient minimization. 
Here $\gamma$ is the magnitude of a planar shear deformation, applied following the procedure in Refs.~\cite{ema_mismatch,Majid_criticallength}. The soft modes of $X^{\rm dq}$ are obtained by analyzing the local curvature of the PEL at $\{{\bf r}_i^{\rm dq}\}$ by diagonalizing the Hessian matrix to obtain the $2N$ eigenfrequencies $\{ \omega_j \}$ and the normalized eigenmodes $\{ {\bf e}_{\omega_j} \}$. To characterize the soft mode localization, we calculate the participation fraction of each particle $i$ in each eigenmode ${\bf e}_{\omega_j}$ as $f_i^j=|\vec{e}^i_{\omega_j}|^2$ (where $\vec{e}^i_{\omega_j}$ is the displacement of particle $i$ along the eigenmode ${\bf e}_{\omega_j}$) and sum this over the $N_m$ lowest frequency modes to obtain $p_i=\sum_{j=3}^{N_{\rm m}}{f_i^j}$. Note that we have excluded the two zero-frequency modes (with index $j = 1, 2$) associated with the translational motion of the center of mass.

\begin{figure}
  \centering
  \includegraphics[width=0.45\textwidth]{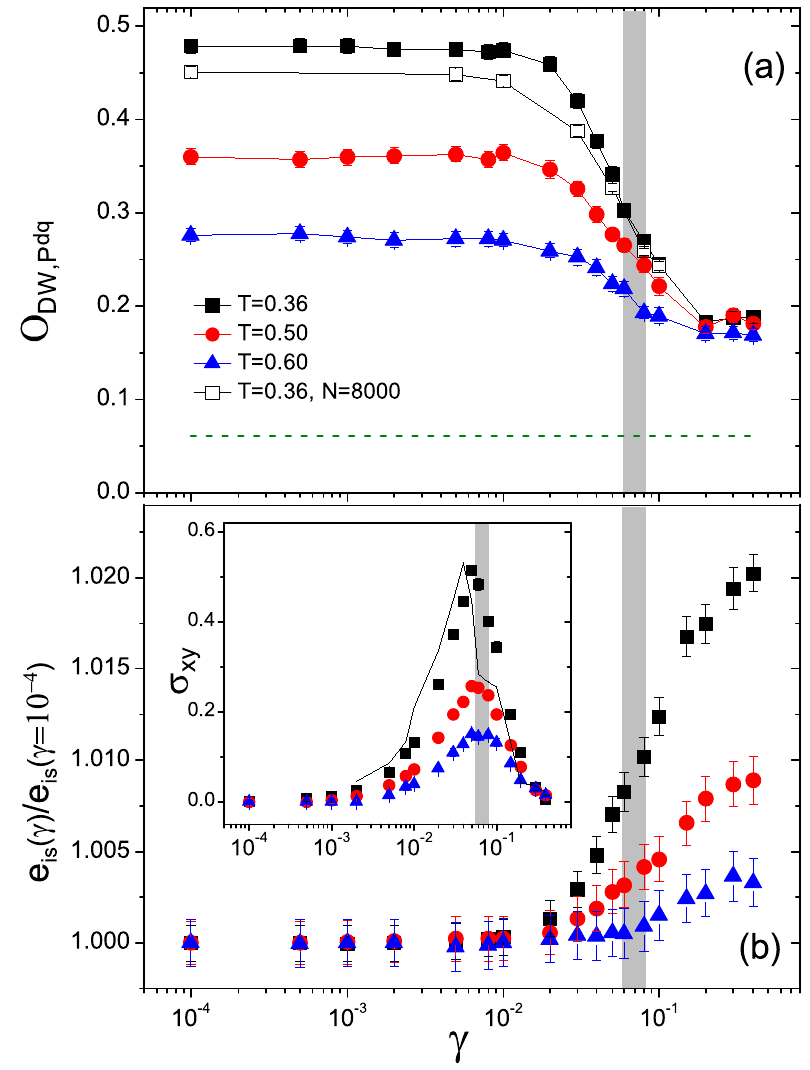}
  \caption{(a) $\OO_{DW,p^{\rm dq}}$  as a function of the shear magnitude $\gamma$. $N=2000$ particles (filled symbols) and $N=8000$ particles (open symbols). Dashed line is $\OO_{DW,{\rm RND}}$, the overlap with a randomly distributed variable, for $N=2000$. (b) The normalized average IS energy $e_{\rm IS}$  as a function of $\gamma$. Inset: average shear stress $\sigma_{xy}^{\rm dq}$ for $X^{\mathrm{dq}}$ as a function of $\gamma$. Solid line is $\frac {d e_{\rm IS}}{d\gamma}$ at $T=0.36$. 
}
\label{overlap_gamma}
\end{figure}

Fig. \ref{overlap_gamma}(a) shows the magnitude of the statistical overlap $\OO_{DW,p^{\rm dq}}$ between the DW-factor of the initial configuration $X$ and the soft mode localization in the IS of the deformed configuration $X^{\mathrm{dq}}$ as a function of  $\gamma$. Throughout the paper, for properties $A$ and $B$ belonging to the same configuration, we define their overlap as $\OO_{A,B}=2(N_A + N_B)^{-1}\sum_{j=1}^N\Theta(A_j)\Theta(B_j)$, where $A_j$ is the value of quantity $A$ for particle $j$, $N_{A}=\sum_j{\Theta(A_j)}$ and $N_{B}=\sum_j{\Theta(B_j)}$. $\Theta(A_j)=1$ if the value of $A_j$ is within the top $25\%$ and belongs to a cluster of more than two particles that satisfy this criterion, otherwise $\Theta(A_j)=0$ \cite{asaphSoftModes}. Error bars are given by the standard deviation over our 50 independent samples.
In Fig.~\ref{overlap_gamma}(a), for small $\gamma$, the overlap $\OO_{DW,p^{\rm dq}}$ is essentially independent of $\gamma$ and increases as the temperature is lowered. However, the overlap drops as $\gamma$ is increased beyond $\gamma_c \approx 0.071 \pm 0.024$. This value appears to be independent of temperature and system size, indicating that, over the entire landscape influenced regime of temperature, the spatial distribution of structurally soft regions does not change significantly until $\gamma_c$.  This transition is accompanied by other changes in the properties of $X^{\mathrm{dq}}$ (see Fig. \ref{overlap_gamma}(b)). For small $\gamma$, the IS energy $e_{\rm IS}(\gamma)$ is approximately constant but starts to increase around the same value of $\gamma$ where the overlap $\OO_{DW,p^{\rm dq}}$ decreases. This shows that deformation magnitudes of the order of $\gamma_c$ push the system into local minima that are energetically less favorable than those sampled by thermal fluctuations. We also find that the increase in the IS energy is accompanied by a rapid increase in the average shear stress $\sigma_{xy}^{\rm dq}(\gamma)$ of $X^{\rm dq}$ (see inset), which reaches a maximum around $\gamma_c$, before dropping back to a value close to zero 
(see Appendix \ref{stress}).%\cite{supporting}. 
These results support the conclusion that $\gamma_c$ is a characteristic property of the PEL.

The change in the spatial distribution of soft regions that is captured by $\gamma_c$ provides a new perspective on the 
hierarchical structure of the amorphous PEL.
On the side of dynamics, the observation of frequent recrossings between nearby ISs has led to the concept of MBs, defined as the collection of minima connected by transitions with greater than $50\%$ probability of return \cite{DoliwaHeuer,Heuer_ISreview}.
%involve higher energy barriers and are typically \emph{irreversible} \cite{DoliwaHeuer}. 
Transitions between MBs have since been identified with higher energy barriers and local \emph{irreversible} cage-breaking events \cite{Heuer_ISreview,Wales01,Wales08,Wales09}. 
The spatial pattern of DH, which is correlated with the soft mode localization, has also been shown to change upon transitions between MBs \cite{Wales01,Appignanesi,asaphSoftModes}.  
Hence, the change in the spatial distribution of soft modes that we detect for $\gamma > \gamma_c$  could be related to MB transitions.
Indeed, the value of $\gamma_c$ that we calculate is similar in magnitude to the typical distance found for the separation between MBs \cite{DoliwaHeuer, Ashwin, note3}. 
On the side of mechanics, recent work on sheared amorphous solids
\cite{Fiocco2013} suggests that $\gamma_c$ may also be related to the onset of irreversible mechanical relaxation processes. A transition from reversible particle motion to diffusive behaviour is in fact detected at a critical shear amplitude $~0.07$, very close to $\gamma_c$. Overall, our results suggest that relaxation processes which have a low probability of 
being reversed in supercooled liquids and in sheared glasses may well be associated with a change in the spatial distribution of structurally soft regions.

\begin{figure}
  \centering
    \includegraphics[width=0.45\textwidth]{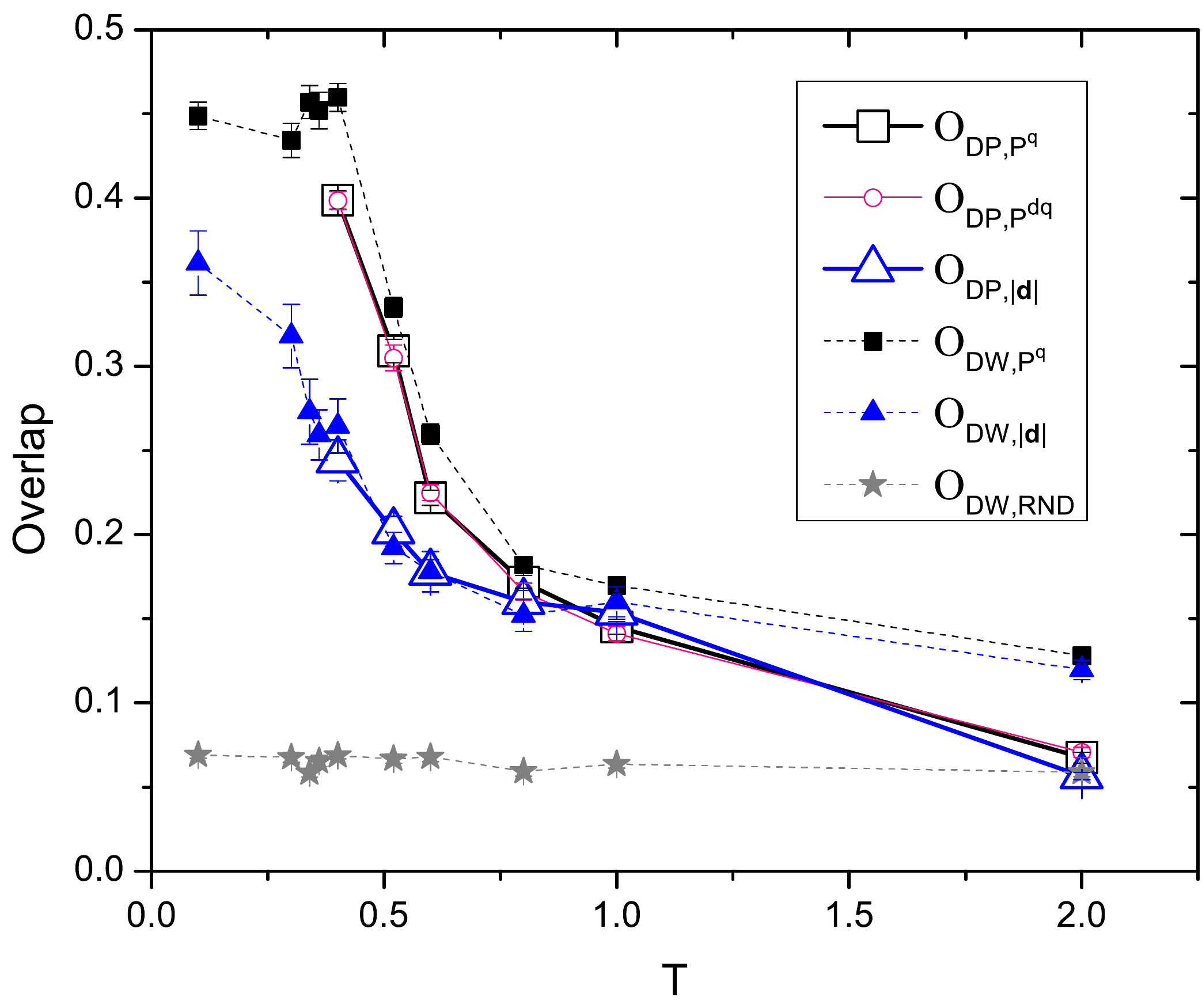}
 \caption{Temperature dependence of the spatial correlation between dynamic heterogeneities and some properties of the underlying potential-energy landscape, as defined in the text. 
 For the non-affine displacements we use $\gamma=10^{-4}$, and for the soft mode localization we use $1\%$ ($N_{\rm m} = 160$) of the lowest-frequency modes. 
}
 \label{overlap_T.fig}
\end{figure}

To better understand the implications of the correlations we detect for $\gamma < \gamma_{c}$ for the supercooled dynamics, at each $T$ we quantify the spatial distribution of DH using the dynamic propensity (DP). The DP is 
defined for each particle $i$ as the iso-configurational average of its squared displacement $DP_{i} = \langle  ({\bf r}_i(t_\alpha)-{\bf r}_i(0))^2 \rangle_{\rm iso}$ over a time interval $t_\alpha$ that probes the $\alpha$-relaxation regime \cite{asaphprl2006}. 
In Fig.~\ref{overlap_T.fig} we plot the overlap between the spatial distribution of the DP in $X$, and the spatial distribution of soft mode localization in $X^{\mathrm{q}}$ and in the nearby IS $X^{\mathrm{dq}}$ for $\gamma = 10^{-4}$, respectively indicated as $\OO_{DP,p^{\rm q}}$ and $\OO_{DP,p^{\rm dq}}$.
%FIGURE 2
%----------------------------- FIG2 ------------------------------------
The data for $\OO_{DP,p^{\rm q}}$ expands on previous results 
\cite{asaphSoftModes,asaphSoftModes2} by showing %for the first time 
that the overlap between the propensity and the soft mode localization increases strongly in the landscape-influenced regime of the dynamics \cite{note2}. The statistical overlap $\OO_{DP,p^{\rm dq}}$ is also very similar to 
$\OO_{DP,p^{\rm q}}$, indicating that the soft mode localization in nearby ISs 
has a similar degree of spatial correlation with DH. The emerging picture is that the soft modes of the initial IS are 
able to predict the spatial pattern of DH occurring over timescales where the system visits many different minima, because of the very similar soft mode pattern of minima belonging to the same meta-basin. 

To extend these observations to lower temperature, we also consider the overlap $\OO_{DW,p^{\rm q}}$,
since the DW-factor is less computationally demanding and 
spatially correlated with the DP \cite{asaphprl2006}. 
$\OO_{DW,p^{\rm q}}$ initially behaves similar to $\OO_{DP,p^{\rm q}}$ but,
at the lowest temperatures where we can equilibrate the system, stops increasing and approaches a plateau. This suggests a crossover to a regime where the structural contribution to DH stops increasing, consistent with recent findings on dynamical correlation lengths \cite{Ludo_nonmonoton}.

When sampling nearby ISs using a shear deformation of amplitude $\gamma$,  one can also identify the regions that undergo non-affine displacements (NAD). The NAD field $\{{\bf d}_i\}$ is defined as the difference between the ISs belonging to the initial configuration $X$ and the final configuration $X^{\rm d}$ when the affine deformation is subtracted, i.e. ${\bf d}_i(\gamma) = {\bf r}_i^{\rm dq}(\gamma) - {\bf r}_i^{\rm q} - \gamma y_i^{\rm q}{\bf e}_x$. In amorphous solids, the NAD field is spatially correlated with the lowest energy soft modes, the evolution of which can predict the emergence of a local plastic instability in the material \cite{lemaitre2007}. In supercooled liquids, the NAD field contains long range correlations that increase in size upon approaching the glass transition \cite{Majid_criticallength,majid_jcp}, however a direct correlation with the soft modes and with the dynamics has not yet been established. 
In Fig. \ref{overlap_T.fig} we show the statistical overlap between the spatial distribution of DH and of the magnitude of the NAD field $\OO_{DW,|{\bf d}|}$ measured for $\gamma =10^{-4}$. This overlap is significantly higher than the one with a randomly distributed variable $\OO_{DW,\rm RND}$ and clearly increases upon entering the landscape-influenced regime, albeit not as strongly as the overlap of the DH with the soft modes. 
This result raises the question of whether there might be a correlation between the soft regions where DH are more likely to appear in the supercooled liquid and the regions prone to plastic events 
once the system has solidified. 
\begin{figure}
  \centering
    \includegraphics[width=0.44\textwidth]{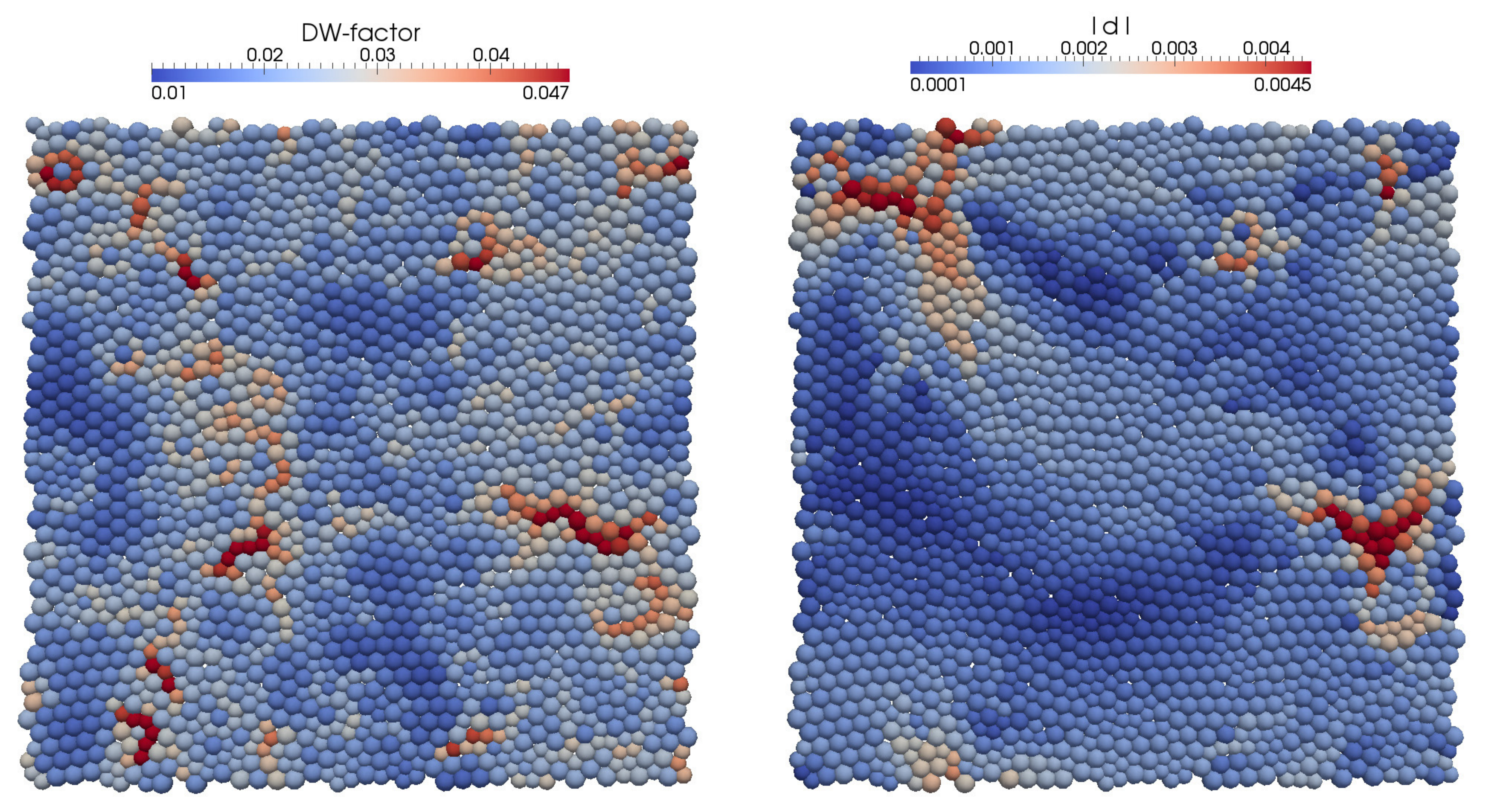}
    \includegraphics[width=0.45\textwidth]{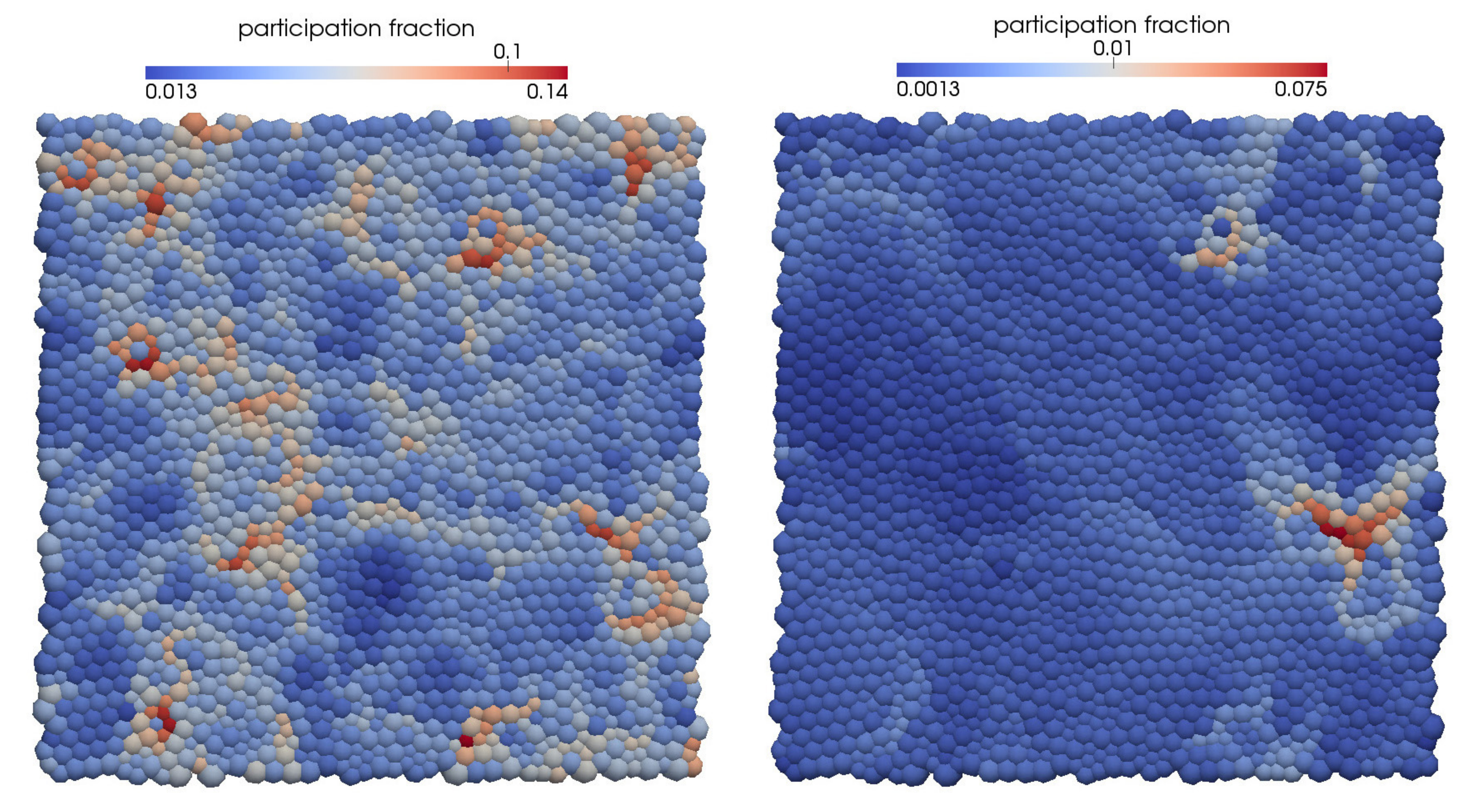}
 \caption{(top) Maps of the local Debye-Waller factor (left) and non-affine displacement (NAD) magnitude (right) for the same initial configuration of $N = 2000$ particles at $T = 0.36$. For NADs we used 
$\gamma=10^{-4}$. (bottom) Maps of the soft mode localization $\{p_{i}\}$ for $X^{\mathrm{q}}$, the initial IS,  
with $N_{\rm m}=60$ (left) and $N_{\rm m}=7$  (right) of the lowest-frequency modes included. 
}
 \label{map.fig}
\end{figure} 

Fig.~\ref{map.fig} shows, for a typical supercooled liquid configuration, the spatial distribution of the DWF (top left); the magnitude of the NAD field $|{\bf d}|$ (top right); the spatial distribution of the soft mode localization in the IS $X^{\rm{q}}$ (lower panels) summed over approx. $1.5\%$ (left) and $0.1\%$ (right) of the lowest-frequency normal modes. We note that at low temperatures the NAD field shows characteristic quadrupolar patterns, typical of NAD in amorphous solids, in agreement with recent findings \cite{Lemaitre-prl2013}. The maps in Fig.~\ref{map.fig} reveal the significant correlations between the DWF, the NAD field, and the soft mode localization that we have discussed so far. Interestingly, they also suggest that DH and NAD are sensitive to different subsets of the soft modes. 
 
\begin{figure}
  \centering
    \includegraphics[width=0.45\textwidth]{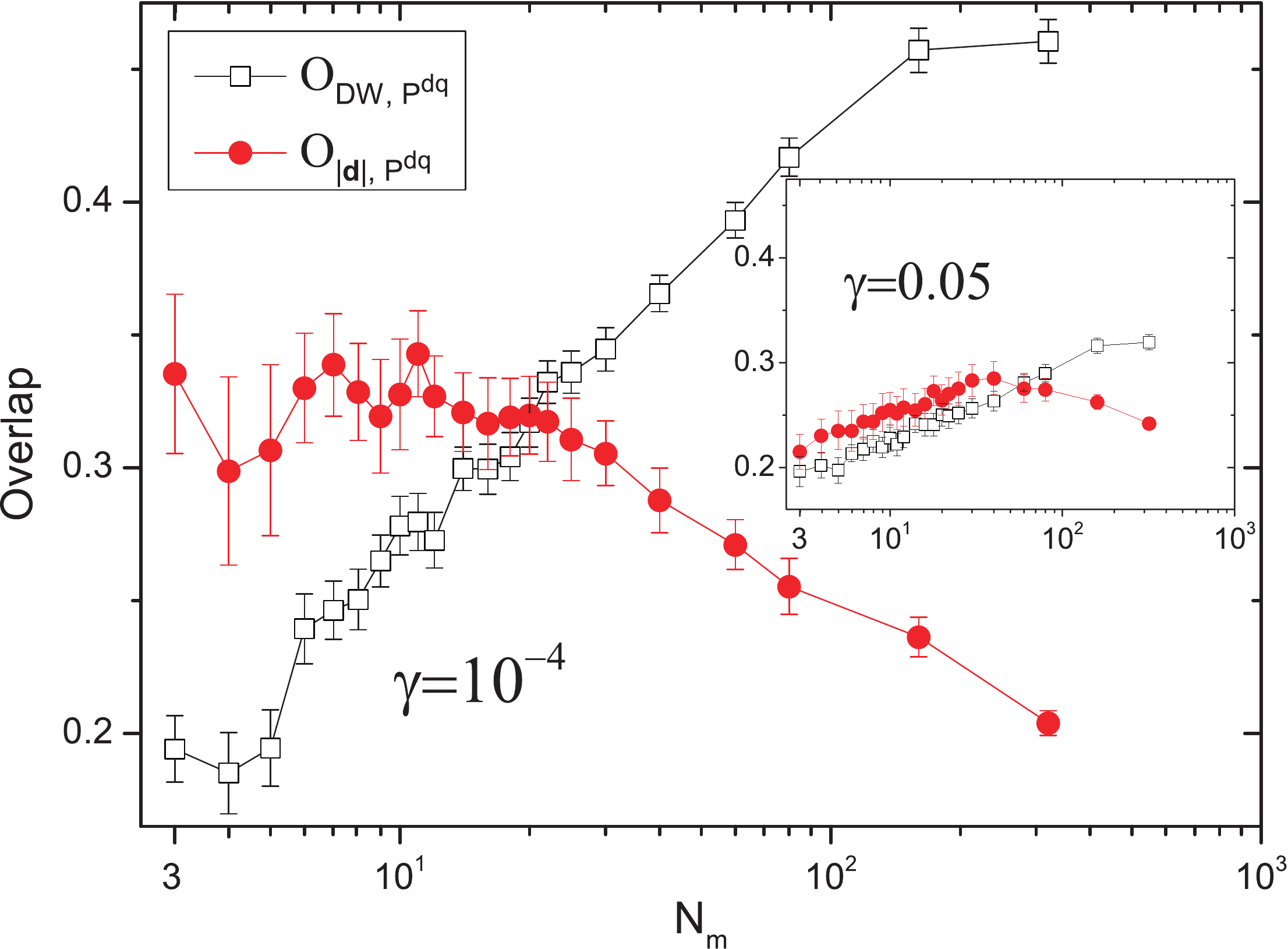}
 \caption{Spatial overlaps between the soft modes of the final IS and, respectively, the DWFs (open) and the NADs (filled), as a function of the number of modes included in the sum over participation fractions, for a system of $N=8000$ particles at $T = 0.40$. Main frame: $\gamma=10^{-4}$; inset: $\gamma=0.05$. 
}
 \label{fig:overlap_Nm}
\end{figure}

We have quantified these differences by computing the overlaps $\OO_{|{\bf d}|,p^{\rm dq}}$ and $\OO_{DW,p^{\rm dq}}$ as a function of the number of low-frequency modes $N_{\rm m}$ included in the participation sum $p^{\rm dq}$ used to determine the soft mode localization. The main part of Fig.~\ref{fig:overlap_Nm} shows these overlaps in the deeply supercooled region ($T = 0.40$) for a small $\gamma=10^{-4}$. We find that the DWF overlap $\OO_{DW,p^{\rm dq}}$ increases with $N_{\rm m}$ until 160 modes are included, whereas the NAD overlap $\OO_{|{\bf d}|,p^{\rm dq}}$ decreases from an initial plateau. That is, about $1\%$ of the lowest-frequency normal modes make a significant contribution to DH, whereas only the very lowest-frequency non-trivial modes have similar spatial structure to the magnitude of the NAD field. 
These results suggest that the regions where structural relaxation is most likely to occur in the supercooled liquid 
can be quite
distinct from the regions in which local plastic instabilities tend to occur in the amorphous solid. 
Interestingly, we also find that around $\gamma_c$ the overlap between NAD and the soft modes of the IS after deformation is no longer sensitive to only the lowest--frequency modes, but instead exhibits a behavior more similar to the DWF overlap, as shown in the inset of Fig.~\ref{fig:overlap_Nm}.
 
%-------------------------CONCLUSIONS------------------------------------------------
In summary, we have provided new insight into the role of the spatial distribution of soft modes for the dynamics of supercooled liquids and explored the link that the modes provide with the amorphous solid eventually formed at the glass transition. 
Our work suggests that the separation between fast reversible and slow irreversible relaxation processes,
usually associated with the hierarchical organization of the PEL, can be related 
to a qualitative change 
in the spatial distribution of soft modes among meta-basins. By investigating non-affinely rearranging regions of 
the IS in the supercooled liquid, we find evidence that the soft modes relevant for DH 
can be quite distinct from the ones 
associated with the mechanical behavior of the solid. Elucidating the nature and origin of this difference can be important to reach deeper understanding and control of the liquid/solid properties of materials close to the glass transition and 
 will be an interesting subject for future work.

% ------------------------ ACKNOWLEDGEMENTS -----------------------------------------
\begin{acknowledgments}
The computational resources of the PolyHub virtual organization are greatly acknowledged.
AW was supported by the Australian Research Council and the Swiss National Science Foundation (Grant No. IZK0Z2\_141601). PI acknowledges support from SNSF (Grant No. 200021\_134626). EDG is supported by the SNSF (Grant No. PP002\_126483/1).
\end{acknowledgments}

\appendix

\section{\label {stress} Stresses in deformed inherent structures}
The relationship between the inherent structure energy $e_{\rm IS}(\gamma)$ and shear stress $\sigma_{xy}^{\rm dq}(\gamma)$ at a given deformation magnitude $\gamma$ observed in Fig.~4  can be explained by a simple analytical argument. 
Consider a system of $N$ particles in a volume $V$ interacting via a central pair potential $\phi(r)$. The potential energy per particle in a given configuration $\{{\bf r}_i\}$ is given by $e=\frac{1}{2N}\sum_{ij}\phi(r_{ij})$. The inherent structure energy is defined as the potential energy in configuration $\{{\bf r}_i^{\rm q}\}$, $e_{\rm IS}=\frac{1}{2N}\sum_{ij}\phi(r_{ij}^{\rm q})$. Consider the IS of the deformed configuration $\{{\bf r}_i^{\rm dq}(\gamma)\}$. For increments $\Delta \gamma$ of the shear magnitude from $\gamma$ to $\gamma+\Delta\gamma$, the IS energy is given by $e_{\rm IS}(\gamma+\Delta\gamma)=\ave{\frac{1}{2N}\sum_{ij}\phi(r_{ij}^{\rm dq}(\gamma+\Delta\gamma))}$. We assume that the particle configuration $\{{\bf r}_i^{\rm dq}(\gamma+\Delta\gamma)\}$ changes only weakly for small $\Delta\gamma$. Then, the inherent structure energy can be expanded to first order in $\Delta\gamma$ giving 
$\frac{de_{\rm iS}}{d\gamma} = v\sigma_{xy}^{\rm dq}+v\varPi_{xy}$, 
where $v=V/N$ and $\sigma_{xy}^{\rm dq}=-\frac{1}{2V}\ave{\sum_{ij}f_{ij,x}^{\rm dq}y_{ij}^{\rm dq}}$ is the average shear stress in the deformed IS configuration $X^{\rm dq}$. Non-affine displacements give rise to additional stress contributions $\varPi_{xy}=\frac{1}{2V}\ave{\sum_{ij}f_{ij,x}^{\rm dq}d_{y,ij}} - \frac{1}{2V}\ave{\sum_{ij}f_{ij,\alpha}^{\rm dq}\delta d_{ij,\alpha}}$, 
where ${\bf d}_{ij}={\bf d}_i-{\bf d}_j$ and $\delta{\bf d}_{ij}$ its derivative with respect to $\gamma$. We define a local strain tensor of particle $i$ in the deformed inherent structure, $\boldsymbol{\varepsilon}_{i\alpha}^{\rm dq}$, such that ${\bf d}_i(\gamma)\approx{\bf d}_j(\gamma)+\boldsymbol{\varepsilon}_{i}^{\rm dq}\cdot{\bf r}_{ij}^{\rm dq}$ holds at least within the interaction range of the potential. 
With this definition, the additional stress can be approximated by 
$\varPi_{xy}\approx
-\ave{\sum_i{\sigma}_{i,x\alpha}^{\rm dq}{\varepsilon}_{i,y\alpha}^{\rm dq}} - \frac{1}{2V}\ave{\sum_{ij}f_{ij,\alpha}^{\rm dq}\delta d_{ij,\alpha}}$. 
The stress tensor for particle $i$ is defined as 
$\sigma_{i,\alpha\beta}^{\rm dq}=-\frac{1}{V}\sum_jf_{ij,\alpha}^{\rm dq}r_{ij,\beta}^{\rm dq}$. 
At zero temperature and in the quasi-static limit, $\varPi_{xy}=0$ and we recover the expression given in Ref.~\cite{Lemaitre06}.  In our case, we find that $\varPi_{xy}$ presents a small correction to $\sigma_{xy}^{\rm dq}$ and therefore the shear stress $\sigma_{xy}^{\rm dq}$ is roughly proportional to the derivative of IS energy with respect to $\gamma$ (the solid line in the inset to Fig. 1b). 

\section{Geometrical picture for NAD}
\begin{figure}
  \centering
  \vspace{0.6cm}
  \includegraphics[width=0.45\textwidth]{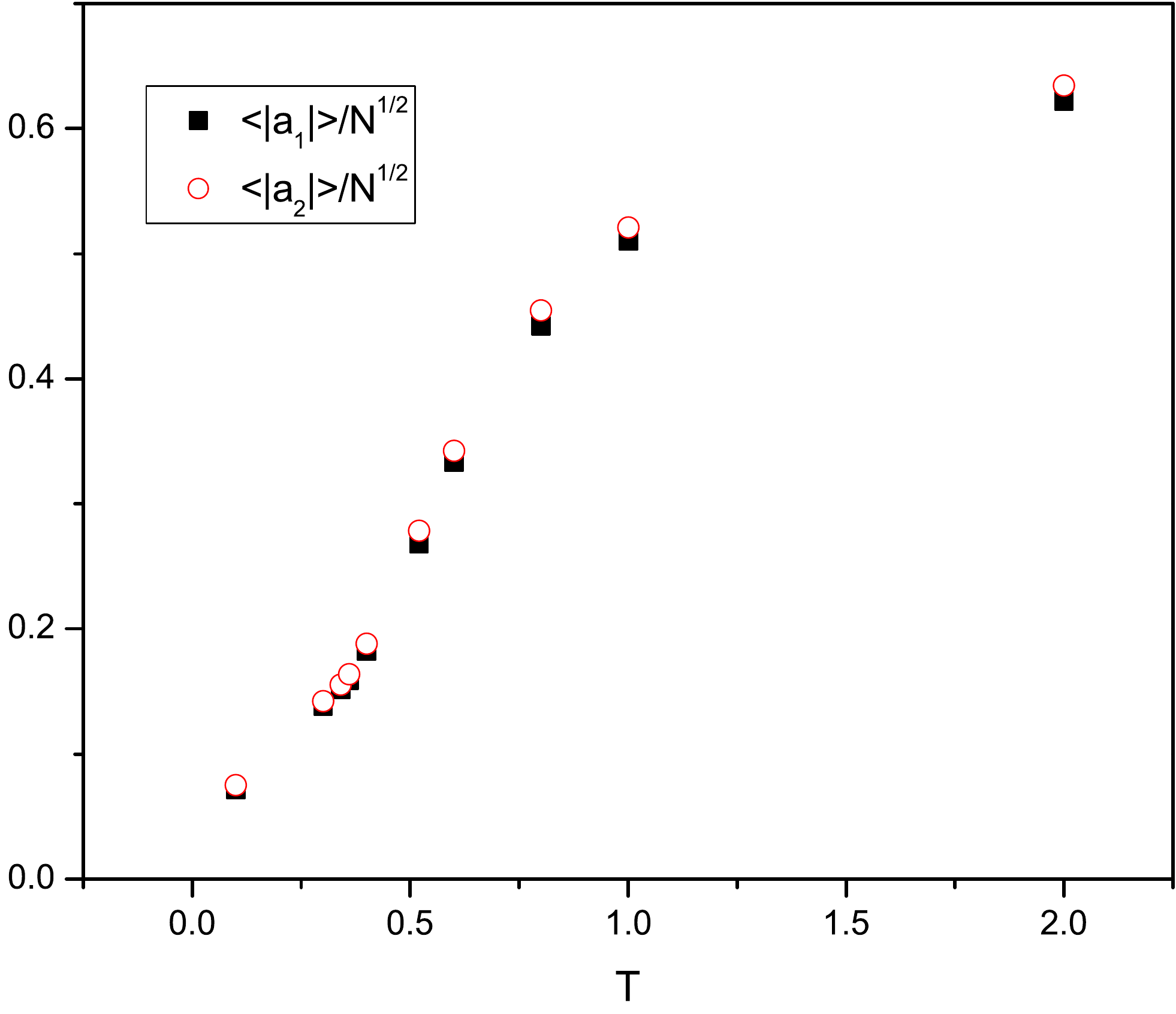}
  \caption{The normalized average magnitude of the quench vectors ${\bf a}_{1}$ and ${\bf a}_{2}$ as a function of temperature ($\gamma=10^{-4}$).  }
    \vspace{0.6cm}
\label{quenches.fig}
\end{figure}

Additional insight into the non-affine displacement (NAD) field 
can be obtained by considering the quenches ${\bf a}_1$ and ${\bf a}_2$ to the inherent structures (ISs) of the initial ($X=\{{\bf r}_i\}$) and deformed ($X^{\rm d}=\{{\bf r}_i^{\rm d}\}$) configurations, respectively. Here, we consider shear deformations of magnitude $\gamma$, ${\bf r}^{\rm d}=(1+\gamma){\bf r}$.  %${\bf r}_i^{\rm d}={\bf r}_i+\gamma y_i{\bf e}_x$. 
In terms of the quench vectors ${\bf a}_{1}={\bf r}^{\rm q}-{\bf r}$ and ${\bf a}_{2}={\bf r}^{\rm dq}-{\bf r}^{\rm d}$, the NAD can be written as ${\bf d}={\bf a}_{2}-(1+\gamma){\bf a}_{1}$. For small deformations, the explicit dependence on the amplitude $\gamma$ drops out and the expression simplifies to ${\bf d} \approx {\bf a}_2 - {\bf a}_1$, which allows to interpret the NAD as the difference of the two quenches. Figure \ref{quenches.fig} shows that the normalized magnitude of the quench vectors $|{\bf a}_1| / N^{1/2}$ and $|{\bf a}_2| / N^{1/2}$ is very similar and decreases with temperature. The normalized magnitude of the quench vectors (also known as ``return distance'' \cite{Stillinger_return}) is a measure for the distance of the initial configuration to its associated inherent structure configuration. Therefore, the decrease of $|{\bf a}_{1}|$ with temperature is expected, since less thermal excitations are available at low temperatures and the configurations are closer to their local minima. Since we consider small shear magnitudes, the mean distance to the local minima is only mildly changed due to the shear deformation.%(see Supplementary Material). 

% ------------------------ REFERENCES -----------------------------------------
%\bibliographystyle{unsrt}
%\bibliography{glass2013}

\begin{thebibliography}{34}
\bibitem [{not({\natexlab{a}})}]{note3}%
  \BibitemOpen
  \href@noop {} {\enquote {\bibinfo {title} {{T}he typical distance found for
  the separation between meta-basins in configuration space in
  {R}ef.~\cite{DoliwaHeuer} was about one particle diameter for a system of
  $65$ particles, corresponding to $\gamma = 1/65 \approx 0.015$.}}\ }
  ({\natexlab{a}})\BibitemShut {NoStop}%
\bibitem [{not({\natexlab{b}})}]{note2}%
  \BibitemOpen
  \href@noop {} {\enquote {\bibinfo {title} {{A}t high temperature the {DWF} no
  longer measures particle localization but our results are robust when
  $t_{DW}$ is changed from $10\tau$ to $\tau$.}}\ }
  ({\natexlab{b}})\BibitemShut {NoStop}%
\end{thebibliography}

%merlin.mbs apsrev4-1.bst 2010-07-25 4.21a (PWD, AO, DPC) hacked
%Control: key (0)
%Control: author (8) initials jnrlst
%Control: editor formatted (1) identically to author
%Control: production of article title (-1) disabled
%Control: page (0) single
%Control: year (1) truncated
%Control: production of eprint (0) enabled
%

\end{document}